\documentclass[11pt]{article}

\usepackage{amsmath,amsfonts,amsbsy,amssymb,array,accents,dsfont}
\usepackage{enumerate,array,latexsym,graphicx,mathrsfs,verbatim,psfrag}
\usepackage{bm} 
\usepackage[normalem]{ulem}
\usepackage{booktabs} 
\usepackage[usenames]{color}
\usepackage[utf8]{inputenc}
\usepackage[english]{babel}
\usepackage{fancybox}

\usepackage{datetime}

\usepackage[nosort]{cite}
\usepackage{chngpage} 

\usepackage{enumitem}
\setlist{itemsep=0pt}

\usepackage[colorlinks=true,      linkcolor=darkblue,      urlcolor=darkblue,      
            filecolor=darkblue,      citecolor=darkblue,       pdfstartview=FitH,     
						pdfpagemode=UseNone,      bookmarksopen=true]{hyperref}  
\usepackage[all]{hypcap}     

\newcommand{\captionfonts}{\small}
\makeatletter  
\long\def\@makecaption#1#2{%
  \vskip\abovecaptionskip
  \sbox\@tempboxa{{\captionfonts #1: #2}}%
 \ifdim \wd\@tempboxa >\hsize
    {\captionfonts #1: #2\par}
  \else
    \hbox to\hsize{\hfil\box\@tempboxa\hfil}%
  \fi
  \vskip\belowcaptionskip}
\makeatother   

\DeclareMathSymbol{\medhatsym}{\mathord}{largesymbols}{"62} 

\DeclareMathSymbol{\medtildesym}{\mathord}{largesymbols}{"65}

\newcommand{\comm}[1]{} 

\def\IR{\mathbb{R}}

\def\({\left(}
\def\){\right)}
\def\[{\left[}
\def\]{\right]}

\def\coeff#1#2{{\textstyle \frac{#1}{#2}}}

\def\One{{\hbox{ 1\kern-.8mm l}}}

\def\barray{\begin{array}}
\def\earray{\end{array}}
\def\be{\begin{equation}}
\def\ee{\end{equation}}
\def\bea{\begin{eqnarray}}
\def\eea{\end{eqnarray}}
\def\bal{\begin{align}}
\def\eal{\end{align}}

\numberwithin{equation}{section} 

\makeatletter
\g@addto@macro\bfseries{\boldmath}
\makeatother

\definecolor{cardinal}{rgb}{0.6,0,0}
\definecolor{darkgreen}{rgb}{0,0.4,0}
\definecolor{purple}{rgb}{0.5, 0, 0.5}
\definecolor{golden}{rgb}{0.92, 0.7, 0}
\definecolor{midnight}{rgb}{0, 0, 0.5}
\definecolor{darkblue}{rgb}{0, 0, 0.8}

\def\coeff#1#2{\relax{\textstyle {#1 \over #2}}\displaystyle}

\def\IR{\mathds{R}}
\def\ZZ{\mathds{Z}}

\def\cA{{\cal A}}

\def\cK{{\cal K}}

\def\cN{{\cal N}}
\def\cO{{\cal O}}

\def\nBPS#1{$\frac{1}{#1}$-BPS}

\def\cO{{\cal O}}

\begin{document}


\begin{flushright}
%
%
\end{flushright}

\vspace{14mm}

\begin{center}

{\huge \bf{Tidal Stresses and Energy Gaps  }} \medskip \\
{\huge \bf{in Microstate Geometries}}\medskip


\vspace{13mm}

\centerline{{\bf Alexander Tyukov$^1$, Robert Walker$^{1}$ and Nicholas P. Warner$^{1,2}$}}
\bigskip
\bigskip
\vspace{1mm}

\centerline{$^1$\,Department of Physics and Astronomy,}
\centerline{University of Southern California,} \centerline{Los
Angeles, CA 90089-0484, USA}
\bigskip
\centerline{$^2$\,Department of Mathematics,}
\centerline{University of Southern California,} \centerline{Los
Angeles, CA 90089, USA}

\vspace{4mm}

{\small\upshape\ttfamily  ~tyukov @ usc.edu,  walkerra @ usc.edu, ~warner @ usc.edu} \\

\vspace{10mm}
 
\textsc{Abstract}

\begin{adjustwidth}{17mm}{17mm} 
 %
\vspace{3mm}
\noindent
We compute energy gaps and study infalling massive geodesic probes in the new families of scaling, microstate geometries that have been constructed recently and for which the holographic duals are known.  We find that in the deepest geometries, which have the lowest energy gaps, the geodesic deviation shows that the stress reaches the Planck scale long before the probe reaches the cap of the geometry.  Such probes must therefore undergo a stringy transition as they fall into  microstate geometry.  We discuss the  scales associated with this transition and comment on the implications for scrambling in microstate geometries.
%
\end{adjustwidth}


\end{center}


\thispagestyle{empty}

\newpage


\baselineskip=14pt
\parskip=2pt

\tableofcontents


\baselineskip=15pt
\parskip=3pt

\section{Introduction}
\label{Sect:introduction}
 
One of the most remarkable properties of microstate geometries is that they can approximate the exterior region of a BPS black hole to arbitrary precision. This means that they can have extremely long AdS throats that cap-off arbitrarily close to where the black-hole horizon would be. Such  backgrounds are often referred to as {\it scaling} microstate geometries and the presence of the long AdS throat means that one can use AdS/CFT to determine the dual microstate structure that is being captured by the microstate geometry. Indeed, recent work \cite{Bena:2015bea,Giusto:2015dfa,Bena:2016agb,Bombini:2017got,Bena:2016ypk,Bena:2017fvm,Bena:2017geu,Bena:2017upb} has been very successful in identifying the states of the D1-D5 CFT that can be captured by fluctuations of single-centered, microstate geometries in six dimensions. 

While the supergravity solutions have continuous parameters describing their energy, charges, angular momenta and moduli, one knows that most of these parameters will ultimately need to be quantized.  As a result, the continuous free parameters of the supergravity solution will typically become rational numbers with specific denominators.  Indeed, these considerations became extremely important in limiting the depths of the AdS throats in microstate geometries. If the throats can actually be made arbitrarily deep then there is a paradox \cite{Bena:2007qc}: the energy gap of the theory can be made arbitrarily small, well below that of the dual CFT, and such geometries could store arbitrary amounts of information, exceeding that of the black hole.  In the simplest examples, the quantization of angular momentum led to a limit on the supergravity moduli \cite{Bena:2006kb,Bena:2007qc} and a far more sophisticated, and more general analysis, was obtained in \cite{deBoer:2008zn,deBoer:2009un} by quantizing the moduli space of multi-centered solutions.  This showed that the tuning of the angular positions of each cycle was limited by the quantized  angular momentum carried by that cycle, thereby limiting the classical depth of microstate geometries.

The analysis of \cite{Bena:2006kb,Bena:2007qc,deBoer:2008zn,deBoer:2009un} established two extremely important results about microstate geometries.  First, quantization not only limits the depth of the throat but also enables one to derive the energy gap of supergravity solution by looking at the longest wavelength mode that could reside at the bottom of the deepest possible throat and then computing its energy {\it including the huge redshift} between the bottom and the top of the throat.  It was one of the early triumphs of the microstate geometry program that this supergravity energy gap precisely matched that of the lowest energy states of the dual D1-D5 CFT.  The second result was that, even though the deepest, scaling microstate geometries were macroscopic solutions with low curvatures and well within the range of the supergravity approximation,  once they became sufficiently deep, the  vast spatial volumes near the bottom of such throats could actually have a phase-space volume of less than $\hbar$.  Thus macroscopic regions of space-time, in which the supergravity approximation appears valid, could be wiped out by a single quantum fluctuation.  This is essentially how quantum mechanics put a limit on the depth of the throats in microstate geometries.

The primary purpose of this paper is to use geodesics to probe the scaling microstate geometries constructed in \cite{Bena:2016ypk,Bena:2017geu,Bena:2017upb}, particularly when these geometries are deep enough to access the very-low energy, intrinsically stringy sectors of the dual D1-D5 CFT.  We will analyze the geodesic deviation for classes of radially infalling, massive particles and, rather surprisingly, we find that {\it such a particle must undergo a stringy transition long before it reaches the cap} of a deep, scaling microstate geometry.  There are several ways to express this result: one is simply that the depth of the throat is dramatically limited if one is to keep the stress forces below the Planck scale. More significantly, we find that, for a scaling microstate geometry of maximal depth whose excitations of the cap have the gap energy, $\cO((N_1 N_5)^{-1})$,   the gravitational stress reaches the Planck scale when the probe is encountering a throat depth commensurate with a CFT  energy scale of $\cO((N_1 N_5)^{-1/2})$, which is that of the typical sector of the D1-D5 CFT.

It is a celebrated fact that there is ``no drama at the horizon of a black hole.''  The tidal forces on an infalling observer near the horizon may be made arbitrarily small by making the mass of the black hole arbitrarily large. The appearance of huge tidal forces  in the middle of the AdS throat of a deep scaling microstate geometry may seem hard to reconcile with the idea that the microstate geometry closely approximates the  black-hole geometry until it caps-off just above the horizon.   Indeed, the microstate geometries of \cite{Bena:2016ypk,Bena:2017geu,Bena:2017upb} closely approximate the extremal BTZ geometry, and yet we will show that probes in the former can encounter Planck-scale stress while probes in the latter encounter no drama whatsoever in the same region of the throat.  

What is happening is that, even though the presence of the cap produces extremely small curving effects higher up in the throat,  the ultra-relativistic speeds of particles falling into very deep throats  greatly magnifies these tiny ``curving effects'' to the extent that they rip the particle into its constituent strings before the particle gets to the cap.  This, once again, highlights the fact that having an AdS vacuum at the horizon scale is an extremely fine-tuned effect, and even tiny deviations combined with large blue-shifts can lead to dramatic effects on probes.  This observation underpinned much of the firewall melodrama \cite{Almheiri:2012rt, Mathur:2012jk,Susskind:2012rm,Bena:2012zi,Susskind:2012uw,Avery:2012tf,Avery:2013exa,Almheiri:2013hfa,Verlinde:2013uja,Maldacena:2013xja,Mathur:2013gua}, but here, with microstate geometries, one now has  a real, physical geometry that is an actual  solution to the supergravity equations of motion,  is consistent with string theory, has a precise holographic interpretation in the D1-D5 CFT  and it allows analysis of the stringy transition of matter falling and scrambling into a fuzzball. 

The microstate geometries that we analyze here can be written in a relatively simple form as  a warped fibration of an $S^3$ over a three-dimensional space-time, $\cK$ \cite{Bena:2017upb}.  For a scaling geometry, it is $\cK$ that closely approximates the BTZ geometry while the $S^3$ remains macroscopic and has  non-trivial warp factors and fibration connections.  These six-dimensional geometries have 
a singular limit in which the  $\cK$ becomes precisely an extremal BTZ black hole, the $S^3$ becomes maximally symmetric while the fibration becomes trivial\footnote{It is in this sense that the microstate geometries can be chosen to be arbitrarily close to BTZ $\times S^3$.}.
From this one might be led to believe that $S^3$ is simply some macroscopic auxiliary space and that most of the interesting physics lies in $\cK$.  Indeed, it was noted in  \cite{Bena:2017upb} that the warp factors exhibit only a modest dimple where the microstate structure is localized.  However, we will show stress forces in the sphere directions not only become extremely large in some directions but also rapidly change sign (twice) as the particle crosses the localized microstate structure.  Thus the huge stress forces not only arise from capping off of the BTZ geometry but also as a result of the non-trivial fibration and  the bumps at the bottom of the throat.

There has been a very interesting recent discussion of  possible instabilities of microstate geometries. Much of this discussion also arises from the study of geodesics but focusses on the trapping of particles near evanescent ergospheres \cite{Eperon:2016cdd,Keir:2016azt,Marolf:2016nwu,Eperon:2017bwq}.  These results are extremely interesting and represent ``feature'' as opposed to a ``bug'' in the microstate geometry program.  Microstate geometries are supposed to represent individual microstates of black holes and sufficiently typical microstate geometries should behave much like a black hole in the same way that a sufficient typical microstate of a box of gas should closely resemble the ensemble average.  Thus microstate geometries should trap particles and those particles should ultimately ``scramble'' into the microstate structure. This process will indeed look just like an instability. The only issue with the analyses in \cite{Eperon:2016cdd,Keir:2016azt,Marolf:2016nwu,Eperon:2017bwq} is that the decay/scrambling channels have been greatly limited for practical reasons and such limited decay channels can suggest that the endpoint of the instability could be singular.  String theory, on the other hand, suggests that the instabilities of infalling particles should somehow result in the scrambling of matter into intrinsically stringy states. This paper reveals precisely such an instability, at least for probes that fall from the top of the BTZ throat: such highly energetic matter encounters Planck scale stresses in scaling microstate geometries.
 
In Section \ref{sec:MGs} we will review the microstate geometries of \cite{Bena:2016ypk,Bena:2017geu} but in the formulation given in \cite{Bena:2017upb}. We will work with geometries that are asymptotic to AdS$_3 \times S^3$ since these are much simpler and yet capture  the essential physics that we wish to explore.  We examine the curvatures and scales of these geometries and verify that they remain safely away from the Planck scale throughout the geometry, and thus the supergravity approximation remains valid.  In Section \ref{sec:EnGap} we analyze the energy gap for the scaling microstate geometries.  We note that one could easily  use the red-shift arguments employed in  \cite{Bena:2006kb,Bena:2007qc} but given that the massless scalar wave equation is separable  \cite{Bena:2017upb}, we use the dispersion relation to analyze the energetics of small fluctuations.  This also reveals an interesting insight into how level-matching in the CFT emerges from smoothness conditions at the cap.   Section \ref{sec:Geodesics} contains our analysis of radial geodesics and the geodesic deviation and we make estimates of where stringy transitions will occur for particles falling from the top of the BTZ throat.  Section \ref{sec:Disc} contains our conclusions.

\section{The microstate geometries} 
\label{sec:MGs}

\subsection{The CFT states and dual geometries}
\label{ss:CFTstates}

We are going to focus on probing geometries that are holographically dual to the pure momentum excitations of the D1-D5 system.  The numbers of underlying D1-branes and D5-branes will be denoted by $N_1$ and $N_5$ respectively.  The ground states of the D1-D5 system can be described by partitioning $N= N_1 N_5$ into strands of lengths between $1$ and $N$ and by the spins of each of the strands (see, for example,\cite{Bena:2015bea}).  One can then excite the strands using operators in the CFT and the standard set of \nBPS{8} states are those that remain in the right-moving Ramond ground state but have arbitrary excitations in the left-moving sector.   Here we will work with microstate geometries that are dual to coherent superpositions of states assembled from strands of length $1$ with spins $|00\rangle$ and $|\!+\!+\rangle$, and we only consider the left-moving excitations obtained by acting with $(L_{-1}- J^3_{-1})$ on the $|00\rangle$ strands.  That is we consider the states:
 \begin{equation}
(|\!+\!+\rangle_1)^{N_{++}} \biggl( \frac{1}{n!} \, (L_{-1}- J^3_{-1})^n\, |00\rangle_1\biggr)^{N_{00}}\,,
 \label{DualStates}
\end{equation}
where $N_{++} + N_{00}  =N \equiv N_1 N_5$ and the subscripts on $|\dots\rangle_1$ indicate the strand length.

The microstate geometries dual to such coherent states were first presented in \cite{Bena:2016ypk}.   Apart from the quantum numbers $N_1, N_5$ and $n$, there are two Fourier coefficients, $a$ and $b$, that determined the numbers of $|\!+\!+\rangle_1$ and $|00\rangle_1$ strands, respectively.  In the supergravity theory, the partitioning of the strands emerges as a regularity condition at the D1-D5 locus and takes the form:
\begin{equation} 
\frac{Q_1Q_5}{R_y^2} ~=~ a^2 + \coeff{1}{2}\, b^2 \,, 
\label{strandbudget1}
\end{equation} 
where $Q_I$ are the supergravity charges and $R_y$ is the radius of the common $y$-circle of the D1 and D5 branes.

The supergravity charges are related to the quantized charges via \cite{Bena:2015bea}:
\begin{equation}
Q_1 ~=~  \frac{(2\pi)^4\,N_1\,g_s\,\alpha'^3}{V_4}\,,\quad Q_5 = N_5\,g_s\,\alpha'\,,
\label{Q1Q5_N1N5a}
\end{equation}
where $V_4$ is the volume of $T^4$ in the IIB compactification to six dimensions.  In particular, it is convenient to define $\cN$ via: 
\begin{equation}
\cN ~\equiv~ \frac{N_1 \, N_5\, R_y^2}{Q_1 \, Q_5} ~=~\frac{V_4\, R_y^2}{ (2\pi)^4 \,g_s^2 \,\alpha'^4}~=~\frac{V_4\, R_y^2}{(2\pi)^4 \, \ell_{10}^8} ~=~\frac{{\rm Vol} (T^4) \, R_y^2}{ \ell_{10}^8} \,,
\label{Q1Q5_N1N5b}
\end{equation}
where $\ell_{10}$ is the ten-dimensional Planck length and  $(2 \pi)^7 g_s^2 \alpha'^4  = 16 \pi G_{10} ~\equiv~ (2 \pi)^7 \ell_{10}^8$.    The quantity, ${\rm Vol} (T^4)  \equiv (2\pi)^{-4} \, V_4$, is sometimes introduced \cite{Peet:2000hn} as a ``normalized volume'' that is equal to $1$ when the radii of the circles in the $T^4$ are equal to $1$.

If one has a supergravity momentum charge, $Q_P$, then it is related to the quantized momentum charge (along the $y$-direction) via:
\begin{equation}
N_P ~=~    \cN \, Q_P \,.
\label{QP_NP}
\end{equation}
%

\subsection{The family of metrics}
\label{ss:metrics}

While the microstate geometries dual to coherent superpositions of states of the form (\ref{DualStates}) were first given in \cite{Bena:2016ypk}, the metrics were rewritten in a much more convenient form in  \cite{Bena:2017upb}.  In particular, they were recast in terms of an $S^3$ fibration over a $(2+1)$-dimensional base, $\cK$: 
\begin{eqnarray}
 ds_6^2 &=& \sqrt{Q_1 Q_5} \,  \frac{\Lambda}{F_2(r)} \bigg[ \frac{F_2(r) \, dr^2}{r^2 + a^2}  \,-\, \frac{2\,F_1(r)}{a^2 (2 a^2 + b^2)^2 \;\! R_y^2 }\bigg(dv + \frac{a^2\,(a^4 + (2 a^2 +b^2) r^2)}{F_1(r)} du \bigg)^2 \cr
& & \qquad \quad \qquad \qquad~ +\, \frac{2 \, a^2 \,r^2 \,(r^2 + a^2)\, F_2(r)}{F_1(r) \, R_y^2} \, du^2 \bigg] \label{sixmet} \\
&& ~+\sqrt{Q_1 Q_5} \, \bigg[  \Lambda \, d\theta^2  \,+\,  
\frac{1}{\Lambda} \sin^2 \theta \, \Big(d\varphi_1 -  \frac{a^2}{(2a^2 +b^2)}\frac{\sqrt{2}}{R_y} (du + dv) \Big)^2  \cr
&& \qquad\qquad\quad~~ + \, \frac{F_2(r)}{\Lambda} \cos^2 \theta \, \Big(d\varphi_2 + \, \frac{1}{(2a^2 +b^2)\, F_2(r)}\frac{\sqrt{2}}{R_y}  \left[ a^2 (du - dv) -  b^2 \, F_0 (r) dv \right] \Big)^2  \bigg] \,,
\nonumber
\end{eqnarray}
where the functions, $F_i(r)$, are defined by:
\begin{equation}
\begin{aligned}
F_0(r) ~\equiv~ & 1 - \frac{r^{2n}}{(r^2 +a^2)^{n}}  \,, \qquad F_1(r) ~\equiv~ a^6 - b^2 \, (2 a^2 + b^2) \,r^2 \, F_0(r) \,, \\
F_2(r) ~\equiv~ &  1  - \frac{a^2\, b^2}{(2 a^2 + b^2) } \,\frac{ r^{2n}}{(r^2 +a^2)^{n+1}}   \, ,
\end{aligned}
  \label{Fdefs}
\end{equation}
and the warp factor, $\Lambda$, is defined by: 
\begin{equation}
\Lambda ~\equiv~ \sqrt{ 1 - \frac{a^2\,b^2}{(2 a^2 +b^2)} \, \frac{r^{2n}}{(r^2 +a^2)^{n+1}} \, \sin^2 \theta  } \,.
  \label{Lambdadef1}
\end{equation}

The coordinates, $u$ and $v$, are the standard null coordinates, which are related to the canonical time and spatial coordinates via:
\begin{equation}
  u ~=~  \coeff{1}{\sqrt{2}} (t-y)\,, \qquad v ~=~  \coeff{1}{\sqrt{2}}(t+y) \,, \label{tyuv}
\end{equation}
where $y$ is the coordinate around $S^1$ with
\begin{equation}
  y ~\equiv~  y ~+~ 2\pi  R_y \,. \label{yperiod}
\end{equation}

The parameters, $a$ and $b$, made their original appearance in the microstate geometry as Fourier coefficients of the underlying supertube profile and of a charge-density fluctuation.  The metric only depends on the RMS values of these profiles, and hence only upon $a^2$ and $b^2$.  The quantized angular momenta and momentum charges of this geometry are related to the Fourier coefficients via \cite{Bena:2016ypk}:
\begin{equation}
J ~\equiv~ J_L ~=~ J_R ~=~  \coeff{1}{2}\, \cN \,  a^2  \,, \qquad \qquad N_P   ~=~  \coeff{1}{2}\, \cN \,n \, b^2\,.
\label{JandP}
\end{equation}
The identity $J_L =  J_R \sim   a^2$ reflects the fact that the only source of angular momentum is the $|\!+\!+\rangle$ supertube and that it has a circular profile in an $\IR^2$ of the spatial $\IR^4$ base geometry.   The  excitations are created only on the $|00\rangle$ supertube and so the momentum  is proportional to $n b^2$.  

It is also useful to note that (\ref{strandbudget1}) can be rewritten as
\begin{equation} 
\frac{b^2}{a^2} ~=~ 2\,\bigg(\frac{Q_1Q_5}{a^2\, R_y^2}  - 1\bigg)  ~=~ 2\,\bigg(\frac{N_1 N_5}{\cN \, a^2}  - 1\bigg) ~=~  \frac{N_1 N_5}{J}  - 2 \,.
\label{strandbudget2}
\end{equation} 
We are going to be particularly interested in microstate geometries that closely approximate non-rotating black holes and so they will be ``deep, scaling geometries,'' with $a^2  ~\ll~  b^2$. Such microstate geometries will be characterized by 
\begin{equation}
b^2 ~ \approx ~\frac{2\,Q_1Q_5}{R_y^2}  \,, \qquad \frac{a^2}{b^2} ~\approx~  \frac{J}{N_1 N_5}  \,.
\label{Scaling1}
\end{equation}
Note that $\frac{a^2}{b^2}$ controls the angular momentum of the CFT state compared to the overall central charge of the CFT. 

\subsection{Limits of the metric}
\label{ss:Limits}

The metric, (\ref{sixmet}), is asymptotic to AdS$_3$ $\times S^3$ at infinity.  Indeed, for large $r$ one has:
\begin{align}
 ds_6^2  ~=~ \sqrt{Q_1 Q_5} \,  \bigg( &  \bigg[ \, \frac{d\rho^2}{\rho^2}  ~-~ \rho^2 \, dt^2 ~+~ \rho^2 \, dy^2 \bigg]  \nonumber\\
 &~+~
\bigg[  d\theta^2  +  \sin^2 \theta \, \bigg(d\varphi_1  - \frac{R_y \, a^2 }{Q_1 Q_5}\, dt\bigg)^2 + \cos^2 \theta \,  \bigg(d\varphi_2- \frac{R_y \, a^2 }{Q_1 Q_5}\, dy\bigg)^2\, \bigg] \bigg)  \label{metinf} \,,
\end{align}
where $\rho \equiv (Q_1 Q_5)^{-\frac{1}{2}} \,r$ and we have used (\ref{tyuv}).  

At the other extreme, $r=0$, the metric becomes degenerate for the simple reason that a (boosted) $y$-circle pinches off.  We therefore expand around $r$, retaining only the $r^2$ terms needed to avoid degeneracy.  We find 
\begin{align}
 ds_6^2  ~=~ \sqrt{Q_1 Q_5} \,  \bigg( &  \bigg[ \,d\rho^2 ~-~ \frac{a^4\, R_y^2}{(Q_1 Q_5)^2} \, dt^2 ~+~\rho^2 \, \bigg(\frac{dy}{R_y}+ \frac{R_y \,b^2\, dt}{2\, Q_1 Q_5}   \bigg)^2  \bigg] \notag\\
 &~+~
\bigg[  d\theta^2  +  \sin^2 \theta \, \bigg(d\varphi_1  - \frac{R_y \, a^2 }{Q_1 Q_5}\, dt\bigg)^2 + \cos^2 \theta \,   \bigg(d\varphi_2 - \frac{dy}{R_y} -  \frac{R_y \, b^2 }{2\, Q_1Q_5}\, dt\bigg)^2\, \bigg] \bigg)  \label{metorig} \,,
\end{align}
where $\rho \equiv r/a$.   One should note that because $y/R_y$ has period $2\pi$, there is no conical singularity at $\rho=0$.  Indeed this geometry is simply that of ${\it Mink}^{(1,2)} \times S^3$.  The geometry thus caps off smoothly at $r =0$.

Finally, there is the $a \to 0$ limit, which leads to the extremal BTZ metric with a round $S^3$:
\begin{equation}
 ds_6^2  ~=~ \sqrt{Q_1 Q_5} \,  \bigg(  \bigg[ \, \frac{d\rho^2}{\rho^2}  - \rho^2 \, dt^2 +  \rho^2 \, dy^2 + \frac{n}{R_y^2} \, (dy+dt)^2 \bigg] 
~+~
\big[  d\theta^2  +  \sin^2 \theta \, d\varphi_1^2 + \cos^2 \theta  \, d\varphi_2^2 \, \big] \bigg)  \label{BTZmet} \,,
\end{equation}
where $\rho \equiv (Q_1 Q_5)^{-\frac{1}{2}} \,r$.
 
\subsection{Some comments on curvatures and the supergravity approximation}
\label{ss:curvatures}

For supergravity to be a valid description of a microstate geometry, the scales of the essential geometric features, and particularly the curvature (length) scale, should be much larger than the Planck length.  For the $T^4$ compactification of IIB supergravity, the scale of the $T^4$ must also be larger than the Planck scale: 
\begin{equation}
{\rm Vol} (T^4) ~>~  \ell_{10}^4  \,.
\label{scales1}
\end{equation}
Furthermore, for the six-dimensional supergravity to be a valid description, the six-dimensional geometry, and $R_y$ in particular, should also be larger than the scale of the $T^4$.  

We will typically find that geometric details and tidal stresses can be expressed as a multiple of some combination of $\ell_{10}$  and  ${\rm Vol} (T^4)$ and we will generically refer to these scales as the Planck/compactification scale.  When possible, we will  use (\ref{scales1}) to relate everything to the ten-dimensional Planck scale.

The AdS$_3$ $\times S^3$ metric is, of course, maximally symmetric and the BTZ metric is that of AdS$_3$ divided by a discrete group and so the curvature is that of AdS$_3$.  It follows that the two  limits,  (\ref{metinf})  and  (\ref{BTZmet}), have radii of curvature, $\ell$, given by:
\begin{equation}
\ell^{-2}  ~=~ \frac{1}{\sqrt{Q_1 Q_5} }  ~\sim~ \frac{\sqrt{2}}{b \,R_y}  \,, \qquad a \to 0 \,.
\label{radcurv1}
\end{equation}
Thus, the curvatures vanish uniformly for large $b$.  

The general metric, (\ref{sixmet}), is obviously much more complicated. However, for small  $\frac{a}{b}$, there is a long BTZ throat and the region  around $r=0$ caps off smoothly in ${\it Mink}^{(1,2)} \times S^3$ while the scale of the $S^3$ remains macroscopic.   One therefore expects that the curvatures remain small and the supergravity approximation to remain valid all the way down to $r=0$.  Indeed, we have computed the Riemann invariant, and we find that it is everywhere regular and, to leading order in inverse powers of $b$, it behaves as:
\begin{equation}
R_{\rho \sigma \mu \nu}R^{\rho \sigma \mu \nu}    ~\sim~\frac{P(r, \sin^2 \theta)}{((r^2+ a^2)^2-  a^2 r^2 \sin^2 \theta)^5} \,\frac{1}{b^2 \,R_y^2}   \,.
\label{RiemInv}
\end{equation}
where $P(r, \sin^2 \theta)$ is a complicated polynomial of degree $20$ in $r$ and degree $4$ in $\sin^2 \theta$.  Thus the Riemann invariant of the general metric, (\ref{sixmet}), exhibits the same overall scaling, (\ref{radcurv1}), in $b R_y$ as the AdS$_3$ $\times S^3$  and the BTZ metric. In particular, the curvature invariant becomes vanishingly small for large $b$.

It therefore appears, at least from the perspective of smoothness and curvature, that supergravity remains valid for any value of $\frac{a}{b}$, including the limit $\frac{a}{b} \to 0$.

Perhaps the most important geometric feature is the cap and this becomes manifest at $r \sim a$.  One has, from (\ref{JandP}), that 
\begin{equation}
 a^2    ~=~  \frac{2 \, J }{\cN}~=~    \frac{2\, \ell_{10}^8 \, J  }{ {\rm Vol} (T^4) \, R_y^2 } \,,
\label{ascale}
\end{equation}
where we have used (\ref{Q1Q5_N1N5b}). Thus, for $J \sim \cO(1)$, $a$ is essentially at the Planck/compactification scale.  This seems to be at the edge of the supergravity limit, however, $a$ itself is not a physical scale in the metric.  Indeed  (\ref{sixmet}) implies that the scale of the cap is approximately 
\begin{equation}
(Q_1 Q_5)^{1/4} \int_{r=0}^{r=a} \, \frac{\sqrt{\Lambda} \,  dr}{\sqrt{r^2 + a^2}}     ~\sim~ (Q_1 Q_5)^{1/4} \,,
\label{capscale}
\end{equation}
which is the horizon scale, and so the cap is always macroscopic.  

It is also very instructive to examine the typical size of the fluctuations in the size of the $S^3$ at various points in the BTZ throat and at the cap.  If one sets $r = a^{(1-\alpha)} b^{\alpha}$ for some  $0 < \alpha < 1$ then, for $b \gg a$, one finds that the circumference of a great circle defined by $\theta$ is given by 
\begin{equation}
2\, (Q_1 Q_5)^{1/4} \int_{\theta=0}^{\theta=\pi} d\theta\, \sqrt{\Lambda}      ~\sim~2 \pi  (Q_1 Q_5)^{1/4} \bigg(1 ~-~ \frac{1}{8}\, \bigg(\frac{a^2}{b^2}\bigg)^\alpha\, \bigg) \,,
\label{circum}
\end{equation}
The deviation from the round circumference has magnitude 
\begin{equation}
\frac{\pi }{4}\,  (Q_1 Q_5)^{1/4}\, \bigg(\frac{a^2}{b^2}\bigg)^\alpha ~\sim~  \frac{\ell_{10}^2 \, J^\alpha  \, (N_1 N_5)^{\frac{1}{4} - \alpha}}{ ({\rm Vol} (T^4))^{\frac{1}{4}}  } \,,
\label{flucscale1}
\end{equation}
This is below the Planck length for $\alpha > \frac{1}{4}$ and macroscopic for $\alpha <  \frac{1}{4}$.  Indeed, for $r= a$ the circle  has circumference
\begin{equation}
2\, (Q_1 Q_5)^{1/4} \int_{\theta=0}^{\theta=\pi} \, \big(1 - 2^{-(n+1)} \sin^2\theta \big)^{1/4} d\theta      \,.
\label{fluccap}
\end{equation}
The deviation is once again of order $\cO((Q_1 Q_5)^{1/4})$ and therefore of similar scale to the horizon of the corresponding black hole. 

In this sense, the metric, (\ref{sixmet}), is extremely close to that of  BTZ $\times S^3$ until close to the bottom of the throat, at which point, the cap and all the essential structural features grow much larger than Planck/compactification scale and indeed have scales comparable to that of the horizon of the corresponding black hole.  One therefore expects that supergravity should accuately capture the essential physical details of this geometry and, most particularly, the cap and the fluctuations caused by the microstate structure.

\section{The energy gap, red shifts and dispersion relations} 
\label{sec:EnGap}

Because of the twisted sectors of the D1-D5 CFT, the excitations of the D1-D5 CFT can have energies that are fractionated relative to the energy scale, $R_y^{-1}$, suggested by the size of the common circle on the D1's and D5's.  Indeed, for strands of length $k$, the lowest energy mode has an energy $\sim k^{-1}$, in units of $R_y^{-1}$.  The maximally-twisted sector has strands of length $N_1 N_5$ and so it gives rise to the lowest energy excitations of the D1-D5 theory, with 
\begin{equation}
E_{gap}    ~=~\frac{1}{N_1 N_5}   \,.
\label{Egap}
\end{equation}
This maximally-twisted sector is extremely important because its extremely small energy gap leads to vast degeneracies of states and these provide a dominant contribution to the entropy of the BPS black hole \cite{Strominger:1996sh}.

\subsection{Red shifts and $E_{gap}$}
\label{ss:RedShift}

One of the triumphs of the microstate geometry program was to show that this sector of the CFT can be accessed using deep, scaling microstate geometries. The original argument was simple \cite{Bena:2006kb,Bena:2007qc}:  scaling geometries could not be made arbitrarily deep because the angular momentum of the geometry was quantized. Setting the angular momentum exactly to zero makes the throat infinitely long, and the geometry becomes singular.  As a result, the lowest angular momentum is $\cO(\hbar)$ and so there is a maximum depth.  The energy gap then emerged from looking at the longest wavelength mode that could reside at the bottom of such a throat and then computing its energy {\it including the huge redshift} between the bottom and the top of the throat.  The result was (\ref{Egap}).  A  much more general argument was given in  \cite{deBoer:2008zn,deBoer:2009un}, where the moduli space of multi-centered microstate geometries was semi-classically quantized. 

If one looks at the coefficient of $dt^2$ in the metric (\ref{metorig}), one immediately notices the huge red-shift factor of $ \frac{a^2\, R_y}{(Q_1 Q_5)^{3/4}}$, where we have taken the square-root to get the scaling of the proper time. One can then use arguments exactly parallel to those of \cite{Bena:2006kb,Bena:2007qc} to arrive at (\ref{Egap}).  However, given some of the remarkable properties of the metric (\ref{sixmet}), we will make a slightly more sophisticated analysis using the dispersion relation for a massless scalar field in  (\ref{sixmet}).

\subsection{Dispersion relations}
\label{ss:Dispersion}

In \cite{Bena:2017upb} it was shown that the massless wave equation was separable in the background (\ref{sixmet}).  In particular, for a  
a generic mode, $\Phi$, of the form  
\begin{equation}
\Phi ~=~ K(r) \, S(\theta) \, e^{i \big(\frac{\sqrt{2}}{R_y}\omega\;\! u  + \frac{\sqrt{2}}{R_y} p \;\! v + q_1\;\! \varphi_1 + q_2\;\! \varphi_2 \big)}\,.
  \label{SepFns}
\end{equation}
the massless wave equation reduces to:
\begin{eqnarray}
 \frac{1}{r} \partial_r \Big(  r(r^2 +a^2) \,  \partial_r K \Big)\,+
\left( \frac{a^2 (\omega+p +q_1)^2}{r^2 + a^2}
-\frac{a^2 (\omega-p-q_2)^2}{r^2}
\right) K \quad &&  \label{SepEqns1}\\
 +\,\, \frac{b^2 \omega  \left(2 a^2 p+ F_0(r) \! \left[2 a^2 (\omega + q_1) + b^2 \omega \right]\right)}{a^2 \left(r^2 + a^2 \right)} K
  &=& \lambda \, K \,, 
	\nonumber \\
\frac{1}{\sin \theta \cos \theta} \partial_\theta \big( \sin \theta \cos \theta \,  \partial_\theta S \big)\,-\, \left(\frac{q_1^2}{\sin^2 \theta} \,+\, \frac{q_2^2}{\cos^2 \theta}\right)\,S     &=& -\lambda \, S \,,
  \label{SepEqns2}
\end{eqnarray}
for some eigenvalue $\lambda$. The second equation has regular solutions based on Jacobi Polynomials of $\cos 2\theta$ provided that 
\begin{equation}
 \lambda ~=~  \ell(\ell+2) \,.
  \label{lambda1}
\end{equation}
Indeed, the solution to the eigenvalue problem, (\ref{SepEqns2}), is simply provided by the harmonic modes on a round $S^3$ of unit radius.

If $b$ were zero then the first eigenvalue problem, (\ref{SepEqns1}), would reduce to finding the harmonic modes on AdS$_3$.  The Laplacian for this is simply:
\begin{equation}
 \frac{1}{r} \partial_r \Big(  r(r^2 +a^2) \,  \partial_r K \Big) ~+~ \left( \frac{a^2 (\omega+p +q_1)^2}{r^2 + a^2}  -\frac{a^2 (\omega-p-q_2)^2}{r^2} \right) K ~=~  \lambda \, K \,.
  \label{LapAdS}
\end{equation}
The  modes of this Laplacian are non-trivial fluctuations of $\rho = r/a$, and hence vary significantly on scales $r \sim a$.  As noted in (\ref{capscale}), such modes actually have a wavelength of order $\cO((Q_1 Q_5)^{1/4})$, which is also the scale of the $S^3$.  The modes we are interested in are precisely these long-wavelength modes, and their counterparts on the $S^3$.  We therefore want  $K(r)$ to be either approximately constant or a slowly varying function of $\rho= r/a$.  

The Laplacian (\ref{LapAdS}) blows up at $r=0$ but this is  a standard artifact of using polar coordinates.  Near $r=0$,  (\ref{SepEqns1})  reduces to:
\begin{equation}
 \frac{1}{r} \partial_r \big(  r\,  \partial_r K \big)     -\frac{ (\omega-p-q_2)^2}{r^2} \, K ~=~  \tilde \lambda \, K \,.
  \label{Lapflat}
\end{equation}
where we have simply put all the constants into a new eigenvalue of $\tilde \lambda$.  This is just  polar form for  the Laplacian on $\IR^2$, as one should expect from the capping off in ${\it Mink}^{(1,2)}$.  Regularity requires that 
\begin{equation}
(\omega-p-q_2) ~=~  m \in \ZZ  \,,
\label{levelmatching}
\end{equation}
and that $K(r) \sim r^m$ as $r \to 0$.  The constant mode, of course, corresponds to $m=0$.

Returning to the original issue, we want to consider modes that are localized around $r=0$ but have the longest wavelength, this means that we  want to solve  (\ref{SepEqns1})  near $r=0$ with $K(r)$ either constant or slowly varying in the cap.  If $K(r)$ is not constant it must satisfy (\ref{levelmatching}) and vanish according to $K(r) \sim r^m$ as $r \to 0$.  We therefore drop the first and third terms in (\ref{SepEqns1}) and set $r=0$ in the remaining part.  Taking $K(r)\sim $ constant ($m=0$) and dropping the overall factor of $K$, one obtains the dispersion relation:
\begin{equation}
 (\omega+p +q_1)^2 ~+~ \frac{2\, b^2}{a^4} \,  \omega  \big( a^2 (p+q_1)+  ( a^2+ \coeff{1}{2} b^2)\,\omega \big)
~=~ \ell(\ell+2) \,,
\label{Dispersion2}
\end{equation}
The simplest non-trivial mode on the $S^3$  is $S(\theta) = \cos 2 \theta$, which has $q_1 = q_2 =0$, $\ell =2$ and hence $\lambda =8$.  We can eliminate $p$ using (\ref{levelmatching}) and we can take $q_1,  q_2$ small and for $\frac{b}{a}$ large, (\ref{Dispersion2}) reduces to
\begin{equation}
\omega  ~=~ \frac{a^2}{b^2}  \,  \mu ~\sim~     \frac{\mu\, J}{N_1 N_5}  \, ,
\label{omres1} 
\end{equation}
where $\mu$ is some number of order $ \sqrt{\ell(\ell+2)} \sim \cO(1)$ and where we have used (\ref{Scaling1}) to arrive at the second identity.  More generally, for $K(r)$ varying slowly in the cap, the constant $\mu$ is modified by the eigenvalue, $\tilde \lambda$, in (\ref{Lapflat}), but this is also a number of order $1$.  Thus we obtain a result of the form (\ref{omres1}) for modes of wavelength of order $\cO((Q_1 Q_5)^{1/4})$ whether they lie on the sphere or in the cap.

The deepest scaling geometries have $J=1$ and so we arrive at 
\begin{equation}
\omega    ~\sim~     \frac{\mu}{N_1 N_5}  \,.
\label{omres2} 
\end{equation}
This implies that for such geometries, the gap energy is given by (\ref{Egap}), as one might have expected.  

However, we have also learnt something else from this analysis of the dispersion relations: Regularity of modes in the microstate geometry imposes level-matching on the dual  CFT state.   Specifically, the quantum numbers $p$ and $\omega$ in (\ref{SepFns}) represent the left-moving and right-moving energies of the modes moving around the $y$-circle.  As we saw in (\ref{levelmatching}), regularity of the modes at $r=0$ requires:
\begin{equation}
(\omega-p)  \in \ZZ  \,.
\label{levelmatching2}
\end{equation}
This is precisely the level-matching condition: $L_0 - \bar L_0 \in \ZZ$ on CFT states.    Our computation, and in particular (\ref{omres2}),  shows the quantum numbers $p$ and $\omega$ can both fractionate in units of $(N_1 N_5)^{-1}$ but their difference must be an integer, exactly as in the dual CFT.   Level-matching is, of course, an expression of world-sheet angular momentum and so its integer quantization is hardly surprising.  However, here we see it emerging from regularity at the cap of a microstate geometry.   

This observation goes some way to explaining why finding fractionated modes in BPS microstate geometries has proven so elusive:  the BPS condition requires the right-moving sector to be in its ground state while regularity of the solution requires level-matching.  This means that smooth BPS microstate geometries cannot see anything other than collective excitations with integer-valued left-moving energies. Energy fractionation is only visible in regular solutions if one looks at non-BPS excitations, like those of (\ref{SepFns}).

It is also interesting to note that this observation meshes  nicely with the results of \cite{Bena:2016agb}, which studied microstate geometries with an orbifold singularity.  These geometries did indeed access fractionated BPS excitations on the left-moving sector but only in combinations that respected level-matching.

\section{Geodesics and probes} 
\label{sec:Geodesics}

\subsection{Radially infalling geodesics}
\label{ss:radial}

Geodesics in the metric  (\ref{sixmet}) were studied in \cite{Bena:2017upb}, where it was shown that there is a conformal Killing tensor and hence there is an additional quadratic integral of the motion for null geodesics.  Here we are going to be concerned with a simple class of {\it time-like} geodesic probes.  For simplicity, we will look at ``equatorial geodesics'' at $\theta = 0$ and $\theta = \frac{\pi}{2}$ and hence $\frac{d\theta}{d \tau} =0$.  One can easily check that such a restriction is consistent with the geodesic equations because of the symmetries of the metric under $\theta \to - \theta$ and $\theta \to \pi - \theta$.  At  $\theta = 0$  the coordinate, $\varphi_1$, degenerates and so we will have $\frac{d\varphi_1}{d \tau} =0$. Similarly, at $\theta = \frac{\pi}{2}$  the coordinate, $\varphi_2$, degenerates and we have $\frac{d\varphi_2}{d \tau} =0$.   

Recall that the warp-factor, $\Lambda$, is given by  (\ref{Lambdadef1}) and so the geodesics with $\theta = \frac{\pi}{2}$ ``see the bump'' more sharply, while the geodesics for $\theta = 0$ see the bump less strongly. 
 
The isometries guarantee the following conserved momenta\footnote{As usual with geodesics, these quantities are ``momenta per unit rest mass,'' and so their dimensions must be adjusted accordingly.}:
\begin{equation}
L_1 ~=~ {K_{(1) \mu }} \frac{dx^\mu}{d \tau} \,,  \qquad L_2 ~=~ {K_{(2) \mu }} \frac{dx^\mu}{d \tau} \,,  \qquad    P ~=~ {K_{(3)   \mu }}  \frac{dx^\mu}{d \tau} \,, \qquad E ~=~ {K_{(4)  \mu }}  \frac{dx^\mu}{d \tau}   \,,
  \label{ConsMom}
\end{equation}
where the $K_{(I)}$  are the Killing vectors: $K_{(J)}  = \frac{\partial}{\partial \varphi_J}$, $K_{(3)}  = \frac{\partial}{\partial v }$ and $K_{(4)}  = \frac{\partial}{\partial u}$.

The standard quadratic conserved quantity coming from the metric is: 
\begin{equation}
g_{\mu \nu} \, \frac{dx^\mu}{d \tau}\,  \frac{dx^\nu}{d \tau}~\equiv~ -1  \,,
  \label{MetInt}
\end{equation}
which means that $\tau$ is the proper time measured on the geodesic. 

One can now use $\frac{d\theta}{d \tau} =0$ and (\ref{ConsMom}) to determine all the velocities with the exception of $\frac{dr}{d \tau}$, however, as usual, this can be determined, up to a sign, from  (\ref{MetInt}).   Since we want to consider infall, we want $\frac{dr}{d \tau} < 0$.

To remove all the centrifugal barriers and enable the geodesic to fall from large values of $r$ down to $r=0$, one must take:
\begin{equation}
L_1 = 0\,, \qquad L_2 = 0 \,, \qquad P ~=~ E  \,.
  \label{CentBarr1}
\end{equation}
One should note that at infinity this means that 
\begin{equation}
\frac{d u}{d\tau}  ~=~ \frac{d v}{d\tau}  ~=~ -\frac{E\sqrt{Q_{1}Q_{5}}}{r^{2}}  \qquad \Rightarrow \qquad \frac{d t}{d\tau}  ~=~  -\frac{E\sqrt{2Q_{1}Q_{5}}}{r^{2}}  \,, \quad\frac{d y}{d\tau}  ~=~ 0  \,.
  \label{velinf1}
\end{equation}
Thus the particle has no $y$-velocity at infinity and, for standard time-orientations ($\frac{d t}{d\tau}>0$), one must have 
\begin{equation}
E  ~<~   0  \,.
  \label{Eneg1}
\end{equation}

Using   (\ref{CentBarr1}), one finds that (\ref{MetInt}) can be reduced to
\begin{equation}
\Big(\frac{dr}{d \tau}\Big)^2  ~=~\frac{(2\, a^2 + b^2 \,F_0(r))}{a^2\, \sqrt{\Lambda}}\,  E^2 ~-~  \frac{(r^2+ a^2)}{R_y\, \sqrt{a^2 + \frac{1}{2} b^2}\, \sqrt{\Lambda}} \,,
  \label{Radvel1}
\end{equation}
where $\Lambda$ is the warp factor, (\ref{Lambdadef1}), and  $F_0(r)$ is defined in (\ref{Fdefs}). Note that for $\theta=0$ one has $\Lambda =1$ while for $\theta=\frac{\pi}{2}$ one has $\Lambda =F_2(r)$. One should also note that in arriving at this expression we have used $\frac{d \theta}{d \tau} =0$ and so it is only valid for geodesics at constant $\theta$.  For infall, one takes the negative square-root.

At $r=0$, this becomes
\begin{equation}
\Big(\frac{dr}{d \tau}\Big)^2  ~=~\frac{2\, Q_1 Q_5}{a^2\,R_y^2}\,  E^2 ~-~  \frac{a^2}{\sqrt{Q_1 Q_5}} \,,
  \label{Radvel2}
\end{equation}
where we have used (\ref{strandbudget1}).  There is thus no centrifugal barrier, and, to leading order, we have:
\begin{equation}
\frac{dr}{d \tau} \Big|_{r=0} ~\sim~\frac{\sqrt{2\, Q_1 Q_5}}{a\,R_y}\,  E  \,.
  \label{Radvel3}
\end{equation}

For $r, b \gg a$, one has:
\begin{equation}
\Big(\frac{dr}{d \tau}\Big)^2  ~=~\Big(2 + \frac{n \,b^2}{r^2}\Big)\,  E^2 ~-~  \frac{ \sqrt{2} \, r^2}{R_y\,b} \,,
  \label{Radvel4}
\end{equation}
which is precisely what one obtains for similar geodesics in the BTZ metric (\ref{BTZmet}) if one uses  (\ref{strandbudget1}) with $b \gg a$.  Note that if these  radial geodesics come to a halt at $r =r_* \gg a$ then 
\begin{equation}
E^2 ~=~ \frac{ \sqrt{2} \, r_*^4}{R_y\,b\, \Big(2\,  r_*^2 +n \,b^2 \Big)} \,.
  \label{Evalue1}
\end{equation}
Also observe that, for $b \gg a$,  the AdS$_3$ region of (\ref{sixmet}) and  (\ref{BTZmet}) starts at around $r  \ge b \sqrt{n}$.

To summarize, the geodesics that we will study are those with $\theta= 0$ or $\theta= \frac{\pi}{2}$, $\frac{d\theta}{d \tau} =0$ and either  $\frac{d\varphi_1}{d \tau} =0$ or $\frac{d\varphi_2}{d \tau} =0$, respectively.  The conserved momenta are restricted to $L_1 =L_2 =0$, $P = E < 0$ while $\frac{dr}{d \tau}$ given by the negative square root in (\ref{Radvel1}). They will start in the  asymptotic AdS$_3$ region, that is, they will have $r_* > b \sqrt{n}$  and, by construction, they will fall all the way to $r =0$.  It is evident from (\ref{Radvel2}) that such a particle will be traveling at a very high speed in the ``Lab Frame''  that is at rest at the bottom of the cap.

For simplicity we will, henceforth, take $n=1$.

\subsection{Tidal forces}
\label{ss:Tides}

For a geodesic with proper velocity, $V^\mu = \frac{dx^\mu}{d \tau}$, the equation of geodesic deviation is:
\begin{equation}
A^\mu ~\equiv~ \frac{D^2 S^\mu}{d \tau^2}  ~=~ - {R^\mu}_{\nu \rho \sigma} \, V^\nu S^\rho   V^\sigma \,,
  \label{Geodev1}
\end{equation}
 where $S^\rho$ is the deviation vector.  By shifting the proper time coordinates of neighboring geodesics one can arrange $S^\rho V_\rho = 0$ over the family of geodesics.  Thus $S^\rho$ is a space-like vector in the rest-frame of the geodesic observer.  One can re-scale $S^\mu$ at any one point so that $S^\mu S_\mu =1 $ and then $A^\mu$ represents the acceleration per unit distance, or the tidal stress.  The skew-symmetry of the Riemann tensor means that $ A^\mu V_\mu  =0$ and so the tidal acceleration is similarly space-like, representing the tidal stress in the rest-frame of the infalling observer with velocity, $V^\mu$.   To find the largest stress one can maximize  the norm, $\sqrt{A^\mu A_\mu}$, of $A^\mu$  over all the choices of $S^\mu$, subject to the constraint $S^\mu S_\mu =1$.  
 
We will consider the geodesics defined in the previous section.  To analyze the stress forces we  introduce what is sometimes called the ``tidal tensor:'' 
\begin{equation}
{\cA^\mu}_\rho ~\equiv~ - {R^\mu}_{\nu \rho \sigma} \, V^\nu \, V^\sigma \,,
  \label{cAdefn}
\end{equation}
and consider its norm and some of its eigenvalues and eigenvectors. In particular, we define
\begin{equation}
|\cA| ~\equiv~  \sqrt{{\cA^\mu}_\rho\, {\cA^\rho}_\mu}  \,.
  \label{cAnorm}
\end{equation}
Note that since $V^\mu = \frac{dx^\mu}{d \tau}$ is dimensionless, $\cA$ has the same dimensions as the curvature tensor, $L^{-2}$.

If $V^\mu$ and $S_{(a)}^\mu$, $a=1, \dots 5$, are orthonormal vectors then it is trivial to see that 
\begin{equation}
|\cA|^2 ~=~ \sum_{a =1}^5 \, {A_{(a)}}^\mu \, {A_{(a)}}{}_\mu\,, 
  \label{cAnormsq}
\end{equation}
where ${A_{(a)}}^\mu$ is given by (\ref{Geodev1}) with $S^\mu = {S_{(a)}}^\mu$.   If there is one dominant direction of maximum stress then one can adapt the basis, $ {S_{(a)}}^\mu$ to this direction and $|\cA|$ will yield this maximum stress.  For our problem,  the maximum stress is spread over multiple directions and so $|\cA|$  will give an estimate of this stress up to a numerical factor of order $1$.  

One should also note that ${\cA^\mu}_\rho$ is not generically symmetric, and so the stress cannot always be directed along the displacement directions, ${S_{(a)}}^\mu$.  

\subsubsection{The scale of the stress and string transitions of probes}
\label{ss:scales}

As a warm-up exercise, we computed ${\cA^\mu}_\rho$ for radially infalling geodesics in the BTZ metric\footnote{Remember we are taking $n=1$.}.  We obtained a result that was independent of the starting point:
\begin{equation}
|\cA|  ~=~ \frac{2}{R_y \, b} ~=~ \frac{\sqrt{2}}{\sqrt{Q_1 Q_5}}  ~=~ \frac{\sqrt{2}}{\sqrt{N_1 N_5}} \,\frac{\sqrt{{\rm Vol} (T^4)}}{\ell_{10}^4} \,.
  \label{cA-BTZ}
\end{equation}
As one would expect, and hope, this becomes vanishingly small compared to the Planck/compactification scale if $N_1$ and $N_5$ are suitably large.  There is, indeed, no tidal drama as one approaches the macroscopic horizon of a black hole. 

The story is very different in the deep, scaling throat of the general metric, (\ref{sixmet}).   We will assume that $b \gg a$ throughout.

First, when an infalling geodesic is near the top of the throat, around $r=b$, one obtains (to leading order in $b$) the same result as for the BTZ black hole, (\ref{cA-BTZ}).  One the other hand, at $r=0$ and $r=a$, one finds: 
\begin{equation}
|\cA|_{r=0}  ~=~  c_1 \,\frac{ E^2 \, b^2}{a^4}  \,, \qquad  \qquad |\cA|_{r=a}  ~=~ c_2  \,\frac{ E^2 \, b^2}{a^4}  \,, 
\label{cAcenter}
\end{equation}
where $E$ is the energy\footnote{As usual with geodesics, $E$ is the energy per unit rest mass, and so is actually dimensionless.} of the geodesic motion and
\begin{equation}
 (c_1,c_2)  ~=~  \bigg(\sqrt{3}\,,\frac{2\sqrt{13}}{9} \, \bigg) \ \ {\rm for} \  \ \theta = \frac{\pi}{2} \,, \\   \qquad(c_1,c_2)  ~=~ \bigg(\sqrt{2}\,,\frac{3}{8} \, \sqrt{\frac{5}{2}} \, \bigg)   \  \ {\rm for}  \  \ \theta = 0\,.
\label{cvals1}
\end{equation}
If the probe is released from rest at $r_* \sim b$, then (\ref{Evalue1}) gives 
\begin{equation}
|E|   ~\sim~  \sqrt{\frac{b}{R_y}}  \,.
  \label{Evalue2}
\end{equation}
(We are dropping numerical factors of order $1$ throughout this discussion.)  Thus we find that, for both classes of geodesic ($\theta = 0,\frac{\pi}{2}$), when they arrive in the vicinity of the cap ($0 \le r \le a$) the stress has a magnitude given by:
\begin{equation}
|\cA|_{cap}  ~\sim~ \frac{ b^3}{a^4 \, R_y}  ~=~ \frac{ b^4}{a^4} \frac{1}{b\, R_y}~\sim~  \frac{(N_1 N_5)^{\frac{3}{2}}}{J^2} \,\frac{\sqrt{{\rm Vol} (T^4)}}{\ell_{10}^4} \,.
  \label{cAcap}
\end{equation}
where we have used the second equation in (\ref{Scaling1}).  Note that at the bottom of the deepest scaling throats, with $J=1$, this stress is super-Planckian. {\it Such an infalling probe must become intrinsically stringy long before it hits the cap!}  Indeed, the only way to avoid the  stringy dissolution of the probe is if the throat is relatively shallow: $J \sim (N_1 N_5)^{\beta}$ for $\beta > \frac{3}{4}$.  From (\ref{omres1}), this corresponds to an energy gap of
\begin{equation} 
E_{gap} ~\sim~\frac{1}{ (N_1 N_5)^{1-\beta} } \,, \qquad   \beta > \frac{3}{4} \,.
\label{shallowgap}
\end{equation}

To find where the string transition must take place in the throat, we computed $|\cA|$ at weighted geometric averages of $a$ and $b$.  Specifically we found that for $r = a^{(1-\alpha)} b^{\alpha}$, $0 < \alpha <\frac{2}{3}$, the dominant term controlling $|\cA|$ is given by
\begin{equation}
|\cA|_{\rm throat}  ~\sim~ \frac{ a^2 \,b^2 \,E^2 }{r^{6}}  ~\sim~\bigg(\frac{b^2}{a^2}\bigg)^{2-3\alpha}\, \frac{1}{b\, R_y}~\sim~ \frac{\sqrt{{\rm Vol} (T^4)}}{\ell_{10}^4}\, \frac{\left(N_{1}N_{5}\right)^{3/2}}{J^{2}}\, \left(\frac{J}{N_{1}N_{5}}\right)^{3\alpha}\,.
  \label{cAthroat}
\end{equation}
If we have the deepest possible throat, with $J \sim 1$, then the tidal stress  hits and exceeds the Planck scale for $\alpha \le 1/2$.  

This implies the central result of this paper: the probe must undergo a stringy transition  as it approaches $r = \sqrt{a b}$.  

It is interesting to note that the dominant stress term that we have singled out in (\ref{cAthroat}) is proportional to $\frac{a^2}{r^6}$. This term  becomes sub-dominant for $\alpha > \frac{2}{3}$ and vanishes when $a=0$.  Indeed, for $a=0$ the metric (\ref{sixmet}) reduces to the BTZ metric and the stress is bounded well below the Planck/compactification scale, as in   (\ref{cA-BTZ}). Thus the stringy transition of the probe is a feature of having the cap at the bottom of a deep throat:  the stress is induced by the probe hurtling through the cap and having its course reversed by the geometry.

\subsubsection{Stress along the sphere and the ``bump''}
\label{ss:evals}

The stress along the sphere directions exhibits some rather remarkable features that can be illustrated by some of the simpler eigenvectors, $U^\rho$, and eigenvalues, $\lambda$, of $\cA$:
\begin{equation}
{\cA^\mu}_\rho \,U^\rho  ~=~ \lambda \, U^\rho    \,. 
  \label{evals0}
\end{equation}
In particular,  $U = \frac{\partial}{\partial \theta}$ is an eigenvector of $\cA$ along the entire geodesics for both $\theta = 0$ and $\theta = \frac{\pi}{2}$ but the eigenvalues are very different for the two classes of geodesic. For $\theta = 0$ we find 
\begin{equation}
\lambda  ~=~ \frac{a^2 r^2 b^2}{4\, R_y\, (r^2 +a^2)^2 (a^2 + \frac{b^2}{2})^{\frac{3}{2}}}  ~\sim~    \frac{a^2 r^2 }{\sqrt{2}\, R_y\, b \, (r^2 +a^2)^2 }\,,
  \label{evals1}
\end{equation}
for large $b$.  In other words, this stress remains very small, unlike the behavior expected from (\ref{cAcenter}).   For $\theta = 0$ one can also show that $U = \frac{\partial}{\partial \varphi_1}$  is an eigenvector with the same eigenvalue,  (\ref{evals1}), as $U = \frac{\partial}{\partial \theta}$.  At $\theta =0$ the $\varphi_1$ coordinate degenerates in the same way that spherical polars degenerate at the origin of $\IR^2$.  This observation about the  $U = \frac{\partial}{\partial \varphi_1}$ being an eigenvector with the same eigenvalue as  $U = \frac{\partial}{\partial \theta}$ reflects the fact that the eigenspace of  (\ref{evals1}) is two-dimensional, corresponding to the slice of tangent space described by $\frac{\partial}{\partial \theta}$ and $\frac{\partial}{\partial \varphi_1}$ in the neighborhood of $\theta =0$.

We have not be able to find any other eigenvectors of $\cA$ for $\theta = 0$  and general values of $r$, however at $r=0$ we find that 
\begin{equation}
U   ~=~  - \frac{R_y}{\sqrt{2}} \, \frac{\partial}{\partial  u}  ~+~    \frac{R_y}{\sqrt{2}} \,  \frac{\partial}{\partial v} ~+~ \, \frac{\partial}{\partial \varphi_2}  \,,
  \label{evecs1}
\end{equation}
is an eigenvector with eigenvalue
\begin{equation}
\lambda  ~=~ -\left(\frac{E^2 b^2}{a^4}  ~+~    \frac{1}{R_y\,\sqrt{(a^2 + \frac{b^2}{2})} }\right)\,,
  \label{evals2}
\end{equation}
which exhibits the growth anticipated by (\ref{cAcenter}).

The picture for $\theta = \frac{\pi}{2}$ is very different and has a rather more interesting structure. First, $U_{(0)} = \frac{\partial}{\partial \theta}$,  $U_{(1)} = \frac{\partial}{\partial \varphi_1}$ and $U_{(2)} = \frac{\partial}{\partial \varphi_2}$ are all eigenvectors of $\cA$, with eigenvalues $\lambda_I$, $I=0,1,2$.   One finds  $\lambda_0 = \lambda_2$ but this is not surprising since $\varphi_2$  degenerates at $\theta = \frac{\pi}{2}$ and we are once again finding a two-dimensional eigenspace in the tangent space at $\theta = \frac{\pi}{2}$.   The general eigenvalues are very complicated but, for $b \gg a$, we have the following limits:
\begin{align}
r \to \infty&: \qquad \lambda_0 = \lambda_2  ~\sim~  \frac{a^2 \, b^2}{4 \, (Q_1 Q_5)^{\frac{3}{2}}} \, \frac{R_y^2}{r^2}\,,    \qquad \lambda_1  ~\sim~ -\frac{a^2 \, b^2}{4 \, (Q_1 Q_5)^{\frac{3}{2}}} \, \frac{R_y^2}{r^2}   \,, \\
r \to a &:\qquad \lambda_0 =  \lambda_2  ~\sim~  \frac{E^2 \, b^2}{9 \, a^4} \,,    \qquad \lambda_1  ~\sim~ - \frac{E^2 \, b^2}{9 \, a^4}     \,, \\
r \to 0&: \qquad \lambda_0 = \lambda_2  ~\sim~  -\frac{E^2 \, b^2}{2 \, a^4} \,,    \qquad \lambda_1  ~\sim~  \frac{E^2 \, b^2}{2 \, a^4}   \,.
  \label{evalsasymp}
\end{align}
 At infinity these eigenvalues fall off more rapidly than anticipated by (\ref{cA-BTZ}) but, in the cap, they exhibit the same scale as indicated by
(\ref{cAcenter}).  Thus, at infinity, the tidal forces along the $S^3$ are rapidly decoupling from the stronger tidal forces of the AdS$_3$.  However, in the cap the tidal forces along the sphere directions are extremely large, exhibiting the typical asymptotic behavior (\ref{cAcenter}).

\begin{figure}
\leftline{\hskip 1.8cm \includegraphics[width=5in]{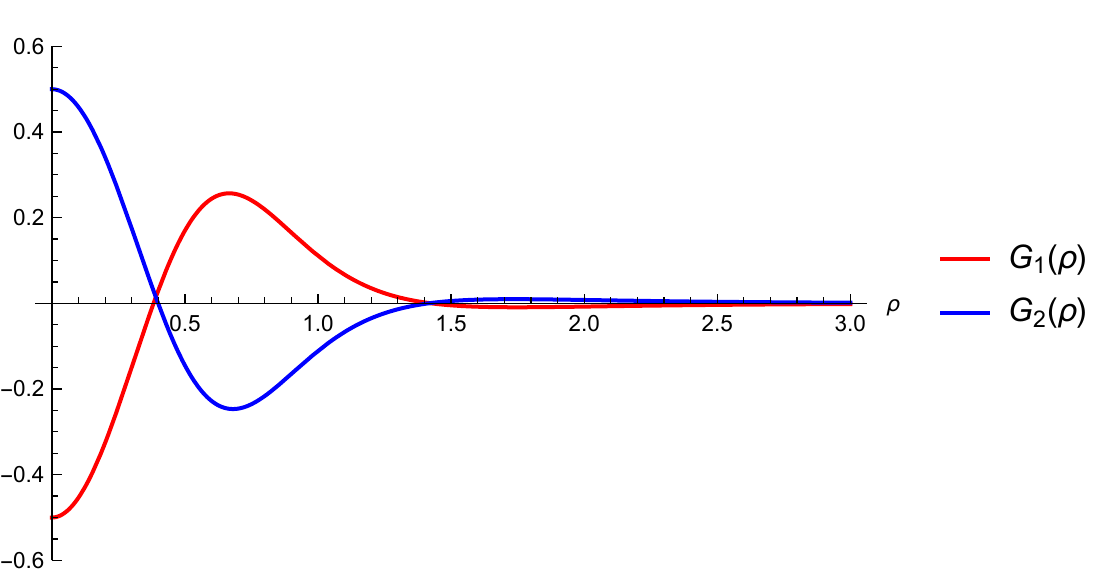}}
\setlength{\unitlength}{0.1\columnwidth}
\caption{\it 
Plot of $G_1(\rho)$ and $G_2(\rho)$ showing the fluctuations of the tidal forces in the $S^3$ directions as the geodesic crosses the cap of the geometry.  As one can see from   (\ref{evals3a}) and  (\ref{evals3b}), the   functions $G_1(\rho)$ and $G_2(\rho)$ take values  $-\frac{1}{2}$ and $+\frac{1}{2}$, respectively, at $\rho =0$. }
\label{fig:Gplots}
\end{figure}

Even more interesting is the leading behavior of these eigenvalues for $b \gg a$.  We find the following leading behavior:
\begin{align}
\lambda_0  ~=~ & \frac{E^2 b^2}{a^4 }  \, G_1(\rho)\,,  \qquad G_1(\rho) ~\equiv~  - \frac{(2 - 11 \rho^2 -18 \rho^4 - 5 \rho^6 + 8 \rho^8)}{4\,(1+\rho^2)(1+\rho^2+\rho^4)^3} \,,
  \label{evals3a} \\ 
  \lambda_1  ~=~ & \frac{E^2 b^2}{a^4 }  \,G_2(\rho)\,,  \qquad G_2(\rho) ~\equiv~   \frac{(2 - 11 \rho^2 -11 \rho^4 +8 \rho^6)}{4\,(1+\rho^2+\rho^4)^3} \,,
  \label{evals3b} 
\end{align}
where $\rho \equiv r/a$.  In particular $G_1$ and $G_2$ both change signs twice on the range $ 0 < r < 2 a$.  (See Fig.~\ref{fig:Gplots}.)  Recall that the microstate structure localizes around $r \sim a$, as is evident from the warp factor, $\Lambda$, in (\ref{Lambdadef1}).  It is evidently the microstate structure that introduces a very bumpy ride for the stresses in the sphere directions.

\subsection{Redshifts and energy scales}
\label{ss:Enthroat}

To estimate the lowest energy excitations of the CFT states that localize in various regions of the throat, we looked, once again at the dispersion relation but now at $r = a^{(1-\alpha)} b^{\alpha}$ with $b\gg a$, $0 < \alpha <\frac{2}{3}$ and $n=1$.

Focussing on the long-wavelength modes that come from excitations on the $S^3$ and taking $K(r) \sim $ constant in the region of interest in the throat, we find 
\begin{equation}
\omega   ~\sim~     \mu\, \bigg(\frac{a^2}{b^2}\bigg)^{1-\alpha}  ~\sim~    \mu\, \bigg(\frac{J}{N_1 N_5}\bigg)^{1-\alpha}  \,.
\label{omres3} 
\end{equation}
where $\mu$ is some number of order $1$.  The deepest throat has $J=1$, and the stringy transition occurs at $\alpha =\frac{1}{2}$.  Thus the energy gap of states that localize near the stringy transition of the probe is:
\begin{equation}
E_{\rm transition}   ~\sim~    \frac{1}{\sqrt{N_1 N_5}} \,.
\label{Etransition} 
\end{equation}

It is interesting to note that while the longest strands have length $N_1 N_5$, in the ensemble of strands it is expected that the typical strand length peaks at $k =  \sqrt{N_1 N_5}$ and  (\ref{Etransition}) is the energy gap for such strands.  Our geodesic probe thus seems to make its stringy transition when it encounters the energy scale appropriate to the most typical strands.

\section{Discussion}
\label{sec:Disc}

As with the older families of scaling, multi-centered microstate geometries \cite{Bena:2006kb,Bena:2007qc,deBoer:2008zn,deBoer:2009un}, we have shown that the quantization of angular momentum in the new microstate geometries of \cite{Bena:2016ypk,Bena:2017geu,Bena:2017upb}  leads  to a limit on the depth of the throat and that excitations at the bottom of this throat have a holographic energy gap that matches the maximally-twisted sector of the dual CFT.  We have also examined the curvatures and the scales of the structural features of the supergravity solutions and found that they all lie well within the range of validity of the supergravity approximation.  One therefore expects that the new microstate geometries represent reliable backgrounds for holographic analyses. 

One of the puzzles in the construction of BPS microstate geometries is finding geometries that correspond to back-reacted, twisted-sector states.  There are some limited examples of such geometries based on orbifolds or fractional spectral flow \cite{Giusto:2012yz, Bena:2016agb}, however it has been very challenging to go beyond these limited constructions and obtain more generic examples.   This has been especially frustrating since it has long been known that BPS supergravity geometries can access states in even the maximally-twisted sector.   In this paper we got some insight into this challenge: regularity of perturbations imposes level-matching on the corresponding states.  For BPS states, the right-moving sector is necessarily in its ground state, which means that the left-moving excitations must have integer energy levels.  Thus BPS microstate geometries will necessarily be dual to states with integer energy levels. Obviously there can be coherent combinations of twisted operators whose energy levels are integral, but such states involve  multiple operator excitations and so may not be easily accessible to the techniques that underlie \cite{Bena:2015bea}, which started from linearized ``seed solutions''  and single operator excitations.  On the other hand, non-BPS excitations, like the scalar fields considered here, are not paired with the right-moving vacuum state and so level-matched states can indeed exhibit highly-fractionated energies.    

This suggests that BPS geometries dual to coherent combinations of twisted operators will necessarily arise through some non-linear supergravity excitation.  One obvious candidate for such geometries are the multi-centered bubbling solutions.  At present the holographic interpretation of bubbling and multi-centered geometries remains, at best, heuristic.  It would be extremely interesting to see if these geometric transitions could be interpreted as some large scale, coherent expression of twisted-sector states. 

The primary impetus behind this work has been to explore geodesic probes of microstate geometries.  We used a special class of probes: massive particles released from rest at $r = r_* \sim b \sim (Q_1 Q_5)^{\frac{1}{4}}$.  In the asymptotically-AdS geometries, these probes start at the top of the BTZ throat.  If one were to add constants to the harmonic functions and obtain asymptotically-flat microstate geometries, these probes would start from the transition region between flat space and the AdS throat.  These probes therefore represent typical infalling particles from near the would-be black hole and, as such, they have extremely high energies compared to the quanta that localize near the cap of the microstate geometry. 

We found that, for deep throats, these probes experience extremely-high stresses from geodesic deviation and that such probes will therefore go through a stringy transition long before they get to the cap. More specifically, we found that to avoid Planck/compactification-scale stress forces on such particles, the depth of the throat must be limited according to: 
\begin{equation}
J \sim (N_1 N_5)^{\beta} \qquad \Leftrightarrow \qquad  E_{gap} ~\sim~\frac{1}{ (N_1 N_5)^{1-\beta} }\,, 
\end{equation}
with $\beta > \frac{3}{4}$.    For the deepest throats, whose states are dual to the maximally-twisted sector, the probe undergoes a stringy transition ``half-way down the throat'' (in the logarithmic sense, $r \sim (a b)^{1/2}$).  The energy scale associated with this transition is
\begin{equation}
  E_{transition} ~\sim~\frac{1}{ \sqrt{N_1 N_5}}\,.
\end{equation}
It is probably not a coincidence that this is also the energy gap of the most typical sector of the D1-D5 system.

These large stress forces are not so much a direct result of the details of the bumps at the bottom of the throat but more a manifestation of the capping off of the throat: They come from the deviation of the microstate geometry from BTZ geometry amplified by the relativistic speed of the infalling particle.  The bumps near the bottom of the microstate geometry can also play a role in the very large tidal stress and we saw that large stress forces  in the sphere directions could even  change sign as the probe encountered the localized microstructure.  

The fact that energetic geodesic probes undergo stringy transitions in the deep, microstate geometries does not invalidate the geometry itself, any more that showers of cosmic rays invalidate the energy levels of the hydrogen atom.  The deep, scaling microstate geometries are still good holographic duals of very low-energy, zero-temperature microstates.  What these probe calculations show is that deep, scaling microstate geometries only make sense in a highly constrained environment and that exposing them to the typical matter of a black-hole environment will generate highly-excited states of the system. In this sense, our geodesic probes reveal an ``instability'' of the BPS microstate geometries, but it is an instability that must be present as part of black-hole physics:  infalling matter must scramble into excited, non-BPS microstates.  

Here we have focussed on geodesic probes and it would be extremely interesting to generalize this work to string and supertube probes and determine exactly how they become excited.  In particular, it would be very interesting to see if one can extend some of the exact CFT techniques developed in \cite{Martinec:2017ztd} to understand the tidal excitations of supertubes.

One of the early, interesting challenges to the fuzzball and microstate-geometry programs was how a perfectly spherical shell of collapsing matter could evolve into a fuzzball rather than merely maintain spherical symmetry and become a Schwarzschild black hole.  The answer lies in the immense density of states in a black hole and the fact that quantum effects become important at the horizon-scale, and so the perfectly spherical shell of matter undergoes a quantum phase transition into fuzzball before a horizon can form \cite{Mathur:2008kg,Mathur:2009zs,Kraus:2015zda, Bena:2015dpt}.   This also means that infall of a probe should also have a description in terms of tunneling into the microstate structure.  This observation underlies the idea of fuzzball complementarity  \cite{Mathur:2010kx, Mathur:2011wg, Mathur:2012jk,
Mathur:2012zp,Mathur:2013gua}.   

In this context, and from the perspective of holography,  BPS microstate geometries represent zero-temperature ground states, or phases, of the underlying D1-D5 system while typical collapse and infall produce extremely high-energy excitations above these ground states.  The results presented in this paper show that if the system lies in a state that is dual to a BPS microstate geometry, then there is a purely classical, tidal phenomenon that will lead to a stringy transition that will ultimately scramble the probe. Thus scrambling could start as a simple, classical phenomenon and may not require one to immediately invoke quantum tunneling into a vast family of degenerate states. On the other hand, fuzzball complementarity suggests that tidal forces in Schwarzschild may be an effective description of the quantum decoherence and scrambling of a probe.  It is therefore quite possible that the tidal effects that we describe in this paper are, in the same spirit, another manifestation of fuzzball complementarity and  quantum scrambling. 


\section*{Acknowledgments}

\vspace{-2mm}
We would like to thank Iosif Bena, David Turton, Stefano Giusto, Masaki Shigemori  and Erik Verlinde for  helpful discussions and  Emil Martinec for  valuable comments on an early draft of this paper. This work  was supported in part by the DOE grant DE-SC0011687.


\begin{adjustwidth}{-1mm}{-1mm} 
\bibliographystyle{utphys}      
\bibliography{microstates}       

\providecommand{\href}[2]{#2}\begingroup\raggedright\begin{thebibliography}{10}

\bibitem{Bena:2015bea}
I.~Bena, S.~Giusto, R.~Russo, M.~Shigemori, and N.~P. Warner, ``{Habemus
  Superstratum! A constructive proof of the existence of superstrata},''
  \href{http://dx.doi.org/10.1007/JHEP05(2015)110}{{\em JHEP} {\bfseries 05}
  (2015) 110},
\href{http://arxiv.org/abs/1503.01463}{{\ttfamily arXiv:1503.01463 [hep-th]}}.

\bibitem{Giusto:2015dfa}
S.~Giusto, E.~Moscato, and R.~Russo, ``{AdS$_{3}$ holography for 1/4 and 1/8
  BPS geometries},'' \href{http://dx.doi.org/10.1007/JHEP11(2015)004}{{\em
  JHEP} {\bfseries 11} (2015) 004},
\href{http://arxiv.org/abs/1507.00945}{{\ttfamily arXiv:1507.00945 [hep-th]}}.

\bibitem{Bena:2016agb}
I.~Bena, E.~Martinec, D.~Turton, and N.~P. Warner, ``{Momentum Fractionation on
  Superstrata},'' \href{http://dx.doi.org/10.1007/JHEP05(2016)064}{{\em JHEP}
  {\bfseries 05} (2016) 064},
\href{http://arxiv.org/abs/1601.05805}{{\ttfamily arXiv:1601.05805 [hep-th]}}.

\bibitem{Bombini:2017got}
A.~Bombini and S.~Giusto, ``{Non-extremal superdescendants of the D1D5 CFT},''
  \href{http://dx.doi.org/10.1007/JHEP10(2017)023}{{\em JHEP} {\bfseries 10}
  (2017) 023},
\href{http://arxiv.org/abs/1706.09761}{{\ttfamily arXiv:1706.09761 [hep-th]}}.

\bibitem{Bena:2016ypk}
I.~Bena, S.~Giusto, E.~J. Martinec, R.~Russo, M.~Shigemori, D.~Turton, and
  N.~P. Warner, ``{Smooth horizonless geometries deep inside the black-hole
  regime},'' \href{http://dx.doi.org/10.1103/PhysRevLett.117.201601}{{\em Phys.
  Rev. Lett.} {\bfseries 117} no.~20, (2016) 201601},
\href{http://arxiv.org/abs/1607.03908}{{\ttfamily arXiv:1607.03908 [hep-th]}}.

\bibitem{Bena:2017fvm}
I.~Bena, P.~Heidmann, and P.~F. Ramirez, ``{A systematic construction of
  microstate geometries with low angular momentum},''
\href{http://arxiv.org/abs/1709.02812}{{\ttfamily arXiv:1709.02812 [hep-th]}}.

\bibitem{Bena:2017geu}
I.~Bena, E.~Martinec, D.~Turton, and N.~P. Warner, ``{M-theory Superstrata and
  the MSW String},'' \href{http://dx.doi.org/10.1007/JHEP06(2017)137}{{\em
  JHEP} {\bfseries 06} (2017) 137},
\href{http://arxiv.org/abs/1703.10171}{{\ttfamily arXiv:1703.10171 [hep-th]}}.

\bibitem{Bena:2017upb}
I.~Bena, D.~Turton, R.~Walker, and N.~P. Warner, ``{Integrability and
  Black-Hole Microstate Geometries},''
\href{http://arxiv.org/abs/1709.01107}{{\ttfamily arXiv:1709.01107 [hep-th]}}.

\bibitem{Bena:2007qc}
I.~Bena, C.-W. Wang, and N.~P. Warner, ``{Plumbing the Abyss: Black Ring
  Microstates},'' \href{http://dx.doi.org/10.1088/1126-6708/2008/07/019}{{\em
  JHEP} {\bfseries 07} (2008) 019},
\href{http://arxiv.org/abs/0706.3786}{{\ttfamily arXiv:0706.3786 [hep-th]}}.

\bibitem{Bena:2006kb}
I.~Bena, C.-W. Wang, and N.~P. Warner, ``{Mergers and Typical Black Hole
  Microstates},'' \href{http://dx.doi.org/10.1088/1126-6708/2006/11/042}{{\em
  JHEP} {\bfseries 11} (2006) 042},
\href{http://arxiv.org/abs/hep-th/0608217}{{\ttfamily arXiv:hep-th/0608217}}.

\bibitem{deBoer:2008zn}
J.~de~Boer, S.~El-Showk, I.~Messamah, and D.~Van~den Bleeken, ``{Quantizing N=2
  Multicenter Solutions},''
  \href{http://dx.doi.org/10.1088/1126-6708/2009/05/002}{{\em JHEP} {\bfseries
  05} (2009) 002},
\href{http://arxiv.org/abs/0807.4556}{{\ttfamily arXiv:0807.4556 [hep-th]}}.

\bibitem{deBoer:2009un}
J.~de~Boer, S.~El-Showk, I.~Messamah, and D.~Van~den Bleeken, ``{A bound on the
  entropy of supergravity?},''
  \href{http://dx.doi.org/10.1007/JHEP02(2010)062}{{\em JHEP} {\bfseries 02}
  (2010) 062},
\href{http://arxiv.org/abs/0906.0011}{{\ttfamily arXiv:0906.0011 [hep-th]}}.

\bibitem{Almheiri:2012rt}
A.~Almheiri, D.~Marolf, J.~Polchinski, and J.~Sully, ``{Black Holes:
  Complementarity or Firewalls?},''
  \href{http://dx.doi.org/10.1007/JHEP02(2013)062}{{\em JHEP} {\bfseries 1302}
  (2013) 062},
\href{http://arxiv.org/abs/1207.3123}{{\ttfamily arXiv:1207.3123 [hep-th]}}.

\bibitem{Mathur:2012jk}
S.~D. Mathur and D.~Turton, ``{Comments on black holes I: The possibility of
  complementarity},'' \href{http://dx.doi.org/10.1007/JHEP01(2014)034}{{\em
  JHEP} {\bfseries 1401} (2014) 034},
\href{http://arxiv.org/abs/1208.2005}{{\ttfamily arXiv:1208.2005 [hep-th]}}.

\bibitem{Susskind:2012rm}
L.~Susskind, ``{Singularities, Firewalls, and Complementarity},''
\href{http://arxiv.org/abs/1208.3445}{{\ttfamily arXiv:1208.3445 [hep-th]}}.

\bibitem{Bena:2012zi}
I.~Bena, A.~Puhm, and B.~Vercnocke, ``{Non-extremal Black Hole Microstates:
  Fuzzballs of Fire or Fuzzballs of Fuzz ?},''
  \href{http://dx.doi.org/10.1007/JHEP12(2012)014}{{\em JHEP} {\bfseries 1212}
  (2012) 014},
\href{http://arxiv.org/abs/1208.3468}{{\ttfamily arXiv:1208.3468 [hep-th]}}.

\bibitem{Susskind:2012uw}
L.~Susskind, ``{The Transfer of Entanglement: The Case for Firewalls},''
\href{http://arxiv.org/abs/1210.2098}{{\ttfamily arXiv:1210.2098 [hep-th]}}.

\bibitem{Avery:2012tf}
S.~G. Avery, B.~D. Chowdhury, and A.~Puhm, ``{Unitarity and fuzzball
  complementarity: 'Alice fuzzes but may not even know it!'},''
  \href{http://dx.doi.org/10.1007/JHEP09(2013)012}{{\em JHEP} {\bfseries 1309}
  (2013) 012},
\href{http://arxiv.org/abs/1210.6996}{{\ttfamily arXiv:1210.6996 [hep-th]}}.

\bibitem{Avery:2013exa}
S.~G. Avery and B.~D. Chowdhury, ``{Firewalls in AdS/CFT},''
\href{http://arxiv.org/abs/1302.5428}{{\ttfamily arXiv:1302.5428 [hep-th]}}.

\bibitem{Almheiri:2013hfa}
A.~Almheiri, D.~Marolf, J.~Polchinski, D.~Stanford, and J.~Sully, ``{An
  Apologia for Firewalls},''
  \href{http://dx.doi.org/10.1007/JHEP09(2013)018}{{\em JHEP} {\bfseries 09}
  (2013) 018},
\href{http://arxiv.org/abs/1304.6483}{{\ttfamily arXiv:1304.6483 [hep-th]}}.

\bibitem{Verlinde:2013uja}
E.~Verlinde and H.~Verlinde, ``{Passing through the Firewall},''
\href{http://arxiv.org/abs/1306.0515}{{\ttfamily arXiv:1306.0515 [hep-th]}}.

\bibitem{Maldacena:2013xja}
J.~Maldacena and L.~Susskind, ``{Cool horizons for entangled black holes},''
  \href{http://dx.doi.org/10.1002/prop.201300020}{{\em Fortsch. Phys.}
  {\bfseries 61} (2013) 781--811},
\href{http://arxiv.org/abs/1306.0533}{{\ttfamily arXiv:1306.0533 [hep-th]}}.

\bibitem{Mathur:2013gua}
S.~D. Mathur and D.~Turton, ``{The flaw in the firewall argument},''
  \href{http://dx.doi.org/10.1016/j.nuclphysb.2014.05.012}{{\em Nucl.Phys.}
  {\bfseries B884} (2014) 566--611},
\href{http://arxiv.org/abs/1306.5488}{{\ttfamily arXiv:1306.5488 [hep-th]}}.

\bibitem{Eperon:2016cdd}
F.~C. Eperon, H.~S. Reall, and J.~E. Santos, ``{Instability of supersymmetric
  microstate geometries},''
  \href{http://dx.doi.org/10.1007/JHEP10(2016)031}{{\em JHEP} {\bfseries 10}
  (2016) 031},
\href{http://arxiv.org/abs/1607.06828}{{\ttfamily arXiv:1607.06828 [hep-th]}}.

\bibitem{Keir:2016azt}
J.~Keir, ``{Wave propagation on microstate geometries},''
\href{http://arxiv.org/abs/1609.01733}{{\ttfamily arXiv:1609.01733 [gr-qc]}}.

\bibitem{Marolf:2016nwu}
D.~Marolf, B.~Michel, and A.~Puhm, ``{A rough end for smooth microstate
  geometries},'' \href{http://dx.doi.org/10.1007/JHEP05(2017)021}{{\em JHEP}
  {\bfseries 05} (2017) 021},
\href{http://arxiv.org/abs/1612.05235}{{\ttfamily arXiv:1612.05235 [hep-th]}}.

\bibitem{Eperon:2017bwq}
F.~C. Eperon, ``{Geodesics in supersymmetric microstate geometries},''
  \href{http://dx.doi.org/10.1088/1361-6382/aa7bfe}{{\em Class. Quant. Grav.}
  {\bfseries 34} no.~16, (2017) 165003},
\href{http://arxiv.org/abs/1702.03975}{{\ttfamily arXiv:1702.03975 [gr-qc]}}.

\bibitem{Peet:2000hn}
A.~W. Peet, \href{http://dx.doi.org/10.1142/9789812799630_0003}{``{TASI
  lectures on black holes in string theory},''} in {\em {Strings, branes and
  gravity. Proceedings, Theoretical Advanced Study Institute, TASI'99, Boulder,
  USA, May 31-June 25, 1999}}, pp.~353--433.
\newblock 2000.
\newblock
\href{http://arxiv.org/abs/hep-th/0008241}{{\ttfamily arXiv:hep-th/0008241
  [hep-th]}}.
\newblock

\bibitem{Strominger:1996sh}
A.~Strominger and C.~Vafa, ``{Microscopic Origin of the Bekenstein-Hawking
  Entropy},'' \href{http://dx.doi.org/10.1016/0370-2693(96)00345-0}{{\em Phys.
  Lett.} {\bfseries B379} (1996) 99--104},
\href{http://arxiv.org/abs/hep-th/9601029}{{\ttfamily arXiv:hep-th/9601029}}.

\bibitem{Giusto:2012yz}
S.~Giusto, O.~Lunin, S.~D. Mathur, and D.~Turton, ``{D1-D5-P microstates at the
  cap},'' \href{http://dx.doi.org/10.1007/JHEP02(2013)050}{{\em JHEP}
  {\bfseries 1302} (2013) 050},
\href{http://arxiv.org/abs/1211.0306}{{\ttfamily arXiv:1211.0306 [hep-th]}}.

\bibitem{Martinec:2017ztd}
E.~J. Martinec and S.~Massai, ``{String Theory of Supertubes},''
\href{http://arxiv.org/abs/1705.10844}{{\ttfamily arXiv:1705.10844 [hep-th]}}.

\bibitem{Mathur:2008kg}
S.~D. Mathur, ``{Tunneling into fuzzball states},''
  \href{http://dx.doi.org/10.1007/s10714-009-0837-3}{{\em Gen. Rel. Grav.}
  {\bfseries 42} (2010) 113--118},
\href{http://arxiv.org/abs/0805.3716}{{\ttfamily arXiv:0805.3716 [hep-th]}}.

\bibitem{Mathur:2009zs}
S.~D. Mathur, ``{How fast can a black hole release its information?},''
  \href{http://dx.doi.org/10.1142/S0218271809016004}{{\em Int. J. Mod. Phys.}
  {\bfseries D18} (2009) 2215--2219},
\href{http://arxiv.org/abs/0905.4483}{{\ttfamily arXiv:0905.4483 [hep-th]}}.

\bibitem{Kraus:2015zda}
P.~Kraus and S.~D. Mathur, ``{Nature abhors a horizon},''
  \href{http://dx.doi.org/10.1142/S0218271815430038}{{\em Int. J. Mod. Phys.}
  {\bfseries D24} no.~12, (2015) 1543003},
\href{http://arxiv.org/abs/1505.05078}{{\ttfamily arXiv:1505.05078 [hep-th]}}.

\bibitem{Bena:2015dpt}
I.~Bena, D.~R. Mayerson, A.~Puhm, and B.~Vercnocke, ``{Tunneling into
  Microstate Geometries: Quantum Effects Stop Gravitational Collapse},''
  \href{http://dx.doi.org/10.1007/JHEP07(2016)031}{{\em JHEP} {\bfseries 07}
  (2016) 031},
\href{http://arxiv.org/abs/1512.05376}{{\ttfamily arXiv:1512.05376 [hep-th]}}.

\bibitem{Mathur:2010kx}
S.~D. Mathur, ``{The information paradox and the infall problem},''
  \href{http://dx.doi.org/10.1088/0264-9381/28/12/125010}{{\em Class. Quant.
  Grav.} {\bfseries 28} (2011) 125010},
\href{http://arxiv.org/abs/1012.2101}{{\ttfamily arXiv:1012.2101 [hep-th]}}.

\bibitem{Mathur:2011wg}
S.~D. Mathur and C.~J. Plumberg, ``{Correlations in Hawking radiation and the
  infall problem},'' \href{http://dx.doi.org/10.1007/JHEP09(2011)093}{{\em
  JHEP} {\bfseries 09} (2011) 093},
\href{http://arxiv.org/abs/1101.4899}{{\ttfamily arXiv:1101.4899 [hep-th]}}.

\bibitem{Mathur:2012zp}
S.~D. Mathur, ``{Black Holes and Beyond},''
  \href{http://dx.doi.org/10.1016/j.aop.2012.05.001}{{\em Annals Phys.}
  {\bfseries 327} (2012) 2760--2793},
\href{http://arxiv.org/abs/1205.0776}{{\ttfamily arXiv:1205.0776 [hep-th]}}.

\end{thebibliography}\endgroup

\end{adjustwidth}


\end{document}